# Near-Infrared Spectral Monitoring of Triton with IRTF/SpeX II: Spatial Distribution and Evolution of Ices

W.M. Grundy[1,2], L.A. Young[1,3], J.A. Stansberry[4],
M.W. Buie[1,3], C.B. Olkin[1,3], and E.F.Young[1,3]

1. Visiting/remote observer at the Infrared Telescope Facility, operated by the University of Hawaii under Cooperative Agreement NCC 5-538 with the National Aeronautics and Space Administration, Science Mission Directorate, Planetary Astronomy Program.
2. Lowell Observatory, 1400 W. Mars Hill Rd., Flagstaff AZ 86001.
3. Southwest Research Institute, 1050 Walnut St., Boulder CO 80302.
4. Steward Observatory, University of Arizona, 933 Cherry Ave., Tucson AZ 87721.

**ABSTRACT**

This report arises from an ongoing program to monitor Neptune's largest moon Triton spectroscopically in the 0.8 to 2.4 µm range using IRTF/SpeX. Our objective is to search for changes on Triton's surface as witnessed by changes in the infrared absorption bands of its surface ices $N_2$, $CH_4$, $H_2O$, CO, and $CO_2$. We have recorded infrared spectra of Triton on 53 nights over the ten apparitions from 2000 through 2009. The data generally confirm our previously reported diurnal spectral variations of the ice absorption bands (Grundy & Young 2004). Nitrogen ice shows a large amplitude variation, with much stronger absorption on Triton's Neptune-facing hemisphere. We present evidence for seasonal evolution of Triton's $N_2$ ice: the 2.15 µm absorption band appears to be diminishing, especially on the Neptune-facing hemisphere. Although it is mostly dissolved in $N_2$ ice, Triton's $CH_4$ ice shows a very different longitudinal variation from the $N_2$ ice, challenging assumptions of how the two ices behave. Unlike Triton's $CH_4$ ice, the CO ice does exhibit longitudinal variation very similar to the $N_2$ ice, implying that CO and $N_2$ condense and sublimate together, maintaining a consistent mixing ratio. Absorptions by $H_2O$ and $CO_2$ ices show negligible variation as Triton rotates, implying very uniform and/or high latitude spatial distributions for those two non-volatile ices.

## 1. Introduction

Despite its small size and frigid surface temperatures, Triton is a spectacularly dynamic world. Its geologically young surface (Croft et al. 1995; Stern & McKinnon 2000; Schenk & Zahnle 2007) is partially coated with a veneer of volatile ices that support and interact with a seasonally-variable atmosphere. Having been imaged up-close by the Voyager 2 spacecraft in 1989, Triton's role is crucial in guiding our thinking about surface-atmosphere interactions on as-yet-unexplored, small, icy planets like Pluto, Eris, and Makemake with similar surface volatile inventories (e.g., Owen et al. 2003; Licandro et al. 2006; Tegler et al. 2008; Abernathy et al. 2009; Merlin et al. 2009). As reviewed by Grundy & Young (2004, hereafter "Paper 1"), Triton's complex seasonal cycle is expected to drive large regional variations in solar heating, leading to dramatic changes in atmospheric pressure and seasonal redistribution of the volatile ices (e.g., Trafton et al. 1998). Observational evidence for ongoing seasonal change on Triton includes its evolving photometric colors at visible and ultraviolet wavelengths (Smith et al. 1989; Buratti et al. 1994; Brown et al. 1995; Young & Stern 2001) and its recently-expanding atmosphere (Elliot et al. 2000). There have also been tentative reports of intermittent or short-term visible wavelength changes (Cruikshank et al. 1979; Buratti et al. 1999; Hicks & Buratti 2004), as well as longer term evolution at near-infrared wavelengths (reviewed by Paper 1).

Reflectance spectroscopy in the near-infrared spectral region (0.7 to 2.5 µm) is especially sensitive to the composition, texture, and distribution of volatile ices on Triton's surface. Observable changes at these wavelengths are expected as ices sublimate or condense. We have been monitoring Triton's near-infrared spectrum for ten years with a single instrument and telescope combination in an effort to detect spectral changes over a variety of timescales. These observations also yield information about the spatial distribution of Triton's ices.

## 2. Observations and Reduction

We successfully observed Triton on 53 nights during 2000-2009 at NASA's Infrared Telescope Facility (IRTF) on Mauna Kea, listed in Table 1. The data were collected using the short cross-dispersed mode of the SpeX spectrograph (Rayner et al. 1998, 2003), which covers wavelengths from 0.8 to 2.4 µm. From fall 2002 onward, most of the observations were done in brief 2 or 3 hour time allocations, taking advantage of the superior scheduling flexibility offered by the IRTF in conjunction with remote observing from the U.S. mainland (e.g., Bus et al. 2002).

**Table 1. Circumstances of Observations**

| UT date of observation mean-time | Sky conditions and H band image size | Sub-Earth longitude (°E) | Sub-Earth latitude (°S) | Phase angle (°) | Triton integration (min) |
|---|---|---|---|---|---|
| 2000/07/22  9:17 | Heavy cirrus, 0.8″ | 344.5 | 50.0 | 0.19 | 9 |
| 2001/07/03 11:56 | Clouds, 2.5″ | 310.3 | 49.9 | 0.87 | 47 |
| 2001/07/04 12:05 | Heavy Cirrus, 1.1″ | 11.9 | 49.9 | 0.84 | 67 |
| 2001/07/05 12:39 | Cirrus, 1.1″ | 74.6 | 49.9 | 0.81 | 48 |
| 2001/07/06 12:03 | Heavy cirrus, 0.9″ | 134.3 | 49.9 | 0.78 | 64 |
| 2001/07/07 12:07 | Cirrus, 1.0″ | 195.7 | 49.9 | 0.75 | 56 |
| 2001/07/08 11:44 | Clear, 1.1″ | 255.9 | 49.9 | 0.72 | 80 |
| 2002/07/15  9:24 | Clear, 0.6″ | 0.7 | 49.8 | 0.58 | 52 |

| UT date of observation mean-time | Sky conditions and H band image size | Sub-Earth | | Phase angle (°) | Triton integration (min) |
|---|---|---|---|---|---|
| | | longitude (°E) | latitude (°S) | | |
| 2002/07/16  9:24 | Scattered clouds, 0.7″ | 61.9 | 49.8 | 0.55 | 54 |
| 2002/07/17  9:53 | Cirrus, 0.8″ | 124.3 | 49.8 | 0.52 | 84 |
| 2002/07/18  9:36 | Thin cirrus, 0.8″ | 184.8 | 49.8 | 0.49 | 64 |
| 2002/07/19  9:31 | Thin cirrus, 0.7″ | 245.8 | 49.8 | 0.46 | 68 |
| 2002/07/20  9:27 | Thin cirrus, 0.8″ | 306.9 | 49.8 | 0.42 | 68 |
| 2002/07/21  9:29 | Thin cirrus, 0.6″ | 8.2 | 49.8 | 0.39 | 68 |
| 2002/07/22  9:21 | Clear, 0.7″ | 69.1 | 49.8 | 0.36 | 84 |
| 2002/08/16 11:10 | Clouds, 1.0″ | 164.1 | 49.9 | 0.47 | 96 |
| 2002/09/16  6:19 | Cirrus, 0.5″ | 249.5 | 49.9 | 1.34 | 84 |
| 2002/10/03  6:32 | Unrecorded, 0.6″ | 211.0 | 49.9 | 1.67 | 48 |
| 2003/07/04 13:58 | Clear, 0.5″ | 101.0 | 49.6 | 0.98 | 64 |
| 2003/07/29 11:59 | Heavy cirrus, 0.7″ | 186.5 | 49.7 | 0.21 | 28 |
| 2003/08/10 10:09 | Some clouds, 0.5″ | 196.4 | 49.7 | 0.19 | 58 |
| 2003/09/10  9:32 | Clear, 0.6″ | 292.6 | 49.8 | 1.13 | 40 |
| 2003/10/16  5:51 | Patchy clouds, 0.5″ | 327.5 | 49.8 | 1.80 | 88 |
| 2004/06/28 13:07 | Cirrus, 0.5″ | 194.9 | 49.3 | 1.19 | 152 |
| 2004/07/28 11:44 | Clear, then fog, 0.6″ | 228.0 | 49.4 | 0.29 | 36 |
| 2004/08/12 12:19 | Clear, 0.7″ | 67.7 | 49.5 | 0.21 | 52 |
| 2004/09/12 10:51 | Clear, 1.0″ | 161.7 | 49.6 | 1.15 | 48 |
| 2004/10/21  6:13 | Partly cloudy, 0.8″ | 18.0 | 49.6 | 1.84 | 52 |
| 2005/07/04 13:06 | Clear, 0.7″ | 244.3 | 48.9 | 1.10 | 146 |
| 2005/08/01 10:22 | Cirrus, 0.6″ | 151.5 | 49.1 | 0.25 | 44 |
| 2005/08/04  9:53 | Cirrus, 0.7″ | 333.9 | 49.1 | 0.15 | 76 |
| 2005/09/19  8:02 | Clear, 0.5″ | 265.2 | 49.3 | 1.26 | 68 |
| 2006/05/26 14:52 | Clear, 0.4″ | 62.3 | 48.4 | 1.86 | 40 |
| 2006/06/26 13:39 | Cirrus, 0.4″ | 157.9 | 48.5 | 1.37 | 128 |
| 2006/07/26 11:01 | Clear, 0.9″ | 188.0 | 48.6 | 0.53 | 72 |
| 2006/08/30  9:14 | Clear, 0.7″ | 166.0 | 48.8 | 0.62 | 48 |
| 2006/10/28  5:59 | Thin cirrus, 0.6″ | 170.3 | 49.0 | 1.85 | 68 |
| 2007/06/21 13:15 | Clear, 0.8″ | 252.6 | 47.9 | 1.53 | 136 |
| 2007/06/22 13:10 | Thin cirrus, 0.6″ | 313.7 | 47.9 | 1.51 | 160 |
| 2007/06/23 13:18 | Clear, 0.6″ | 15.2 | 47.9 | 1.49 | 128 |

| UT date of observation mean-time | Sky conditions and H band image size | Sub-Earth | | Phase angle (°) | Triton integration (min) |
|---|---|---|---|---|---|
| | | longitude (°E) | latitude (°S) | | |
| 2007/06/24 14:06 | Clear, 0.7″ | 78.5 | 47.9 | 1.47 | 88 |
| 2007/06/25 13:13 | Clear, 0.5″ | 137.5 | 47.9 | 1.45 | 132 |
| 2007/06/26 13:09 | Heavy cirrus, 0.7″ | 198.6 | 47.9 | 1.43 | 172 |
| 2007/06/27 13:16 | Humid, 1.2″ | 260.1 | 47.9 | 1.41 | 132 |
| 2007/08/01 11:59 | Clear, 0.9″ | 239.8 | 48.1 | 0.41 | 16 |
| 2008/06/24 14:36 | Clear, 0.5″ | 182.9 | 47.3 | 1.50 | 56 |
| 2008/07/14 10:57 | Clear, 0.5″ | 318.3 | 47.4 | 1.01 | 72 |
| 2008/08/12 12:24 | Clear, 0.9″ | 297.5 | 47.6 | 0.10 | 44 |
| 2008/09/01  7:02 | Clear, 0.6″ | 68.1 | 47.8 | 0.55 | 56 |
| 2008/09/28  9:55 | Partly cloudy, 0.5″ | 288.5 | 48.0 | 1.32 | 60 |
| 2009/06/22 14:35 | Clear, 0.7″ | 102.2 | 46.6 | 1.59 | 32 |
| 2009/06/24 14:25 | Clear, 0.6″ | 224.3 | 46.6 | 1.56 | 80 |
| 2009/07/21 12:07 | Clear, 0.5″ | 71.7 | 46.8 | 0.88 | 64 |

Details of data acquisition and reduction strategies have been described previously (Paper 1; also Grundy et al. 2006). To remove telluric absorptions, instrumental effects, and the solar spectrum, all Triton observations were interspersed between observations of solar analog stars. During 2000-2001, we only used well known solar analogs 16 Cyg B, BS 5968, BS 6060, and SA 112-1333. Unfortunately, Neptune and Triton were far from these stars on the sky plane, resulting in slews large enough to give troublesome instrumental flexure between Triton and star observations. To remedy this, in 2002, we began using the much more nearby solar analog star HD 202282 (spectral class G3V, according to Houk & Smith-Moore 1988) after verifying with SpeX observations that its spectrum matched those of the better known analogs. Neptune's orbital motion forced us to transition to a new nearby solar analog in 2007/2008. We selected BS 8283 for this purpose. Two nights of SpeX observations of both BS 8283 and HD 202282 during 2007 showed that the two stars were spectrally similar at our wavelength range and spectral resolution, despite BS 8283 being an unresolved G0V+G1IV binary system (Neckel 1986; Pourbaix et al. 2004).

Wavelength calibration was derived from observations of SpeX's internal integrating sphere, illuminated by a low pressure argon arc lamp. This calibration was checked against telluric sky emission lines extracted from the Triton frames themselves, which sometimes called for a slight shift relative to the argon spectra, due to flexure within SpeX. Profiles of both arc and sky lines during 2002-2009 (when we used SpeX's 0.3 arcsec slit, in order to maintain as consistent as possible spectral resolution) resembled Gaussians having full width at half maximum (*FWHM*) of 2.5 pixels, implying spectral resolution ($\lambda/FWHM$) between 1600 and 1700. Prior to 2002 we used a wider slit (0.5 arcsec) to admit more light, resulting in measured spectral resolutions in the 1300 to 1400 range. Wavelength uncertainty is less than a pixel in regions having abundant arc and night sky lines. The spectra were not photometrically calibrated, due to variable slit losses from distinct tracking and guiding correction timescales between bright calibration stars and much fainter Triton, along with variable focus and seeing conditions. Also, many of our spectra were collected on non-photometric nights. An average spectrum appears in Fig. 1.

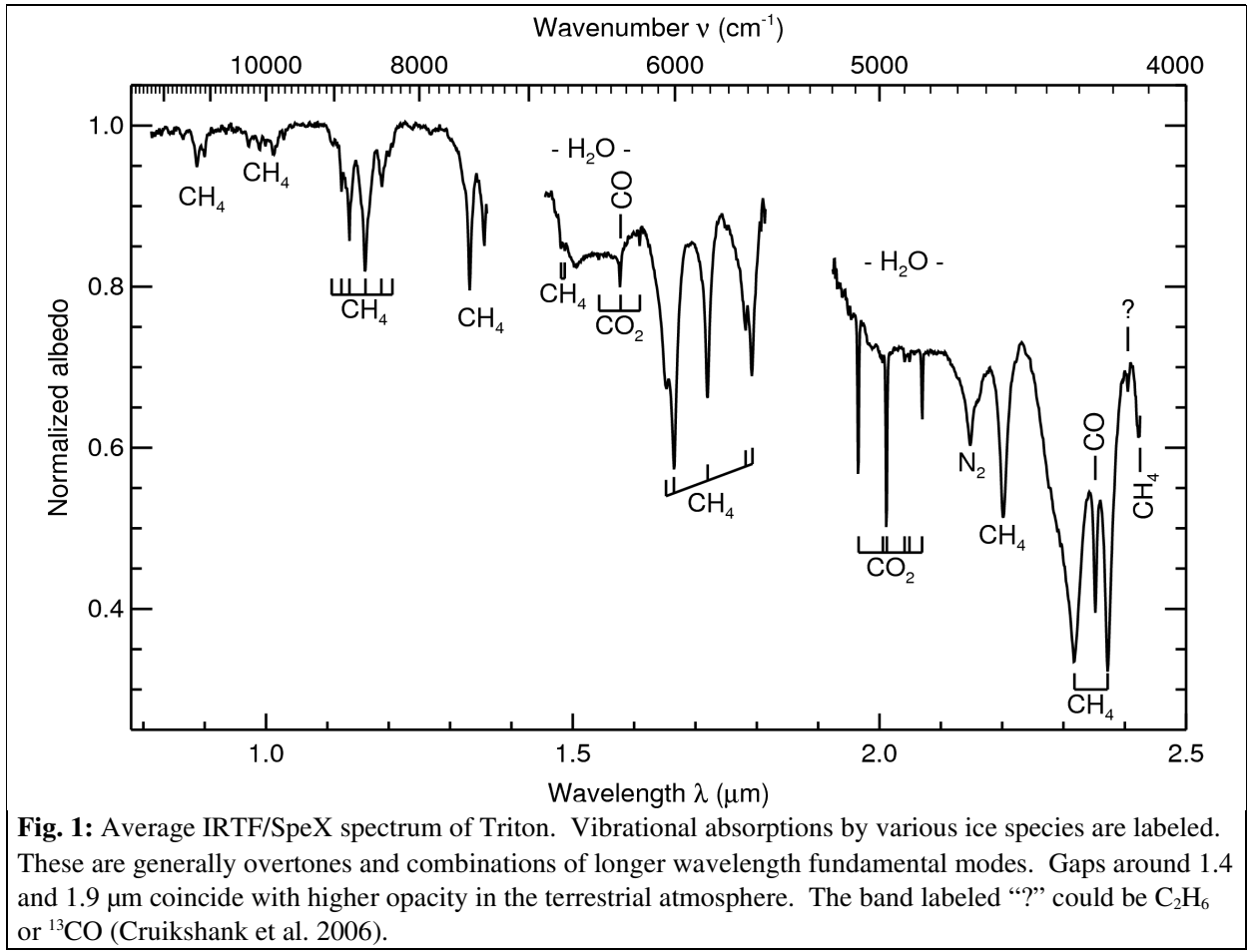

**Fig. 1:** Average IRTF/SpeX spectrum of Triton. Vibrational absorptions by various ice species are labeled. These are generally overtones and combinations of longer wavelength fundamental modes. Gaps around 1.4 and 1.9 μm coincide with higher opacity in the terrestrial atmosphere. The band labeled "?" could be $C_2H_6$ or $^{13}CO$ (Cruikshank et al. 2006).

Examples of individual nightly spectra are shown in Fig. 2 and more can be found in Paper 1.

## 3. Longitudinal Distributions of Ice Species

Based on eight nights of data obtained in 2002, Paper 1 reported diverse patterns of longitudinal variability for Triton's different ice species. The strongest variation, almost a factor of two, was seen in the 2.15 μm $N_2$ ice absorption band, implying a very uneven regional distribution of that species, heavily biased toward low latitudes and the Neptune-facing hemisphere. Regional variations in $N_2$ ice texture could also explain the observed variation, by causing longer mean optical path lengths in those regions. The $CH_4$ ice bands were seen to vary less, and to exhibit maximum absorption at longitudes shifted toward the trailing hemisphere. This observation was puzzling in light of the fact that Triton's $CH_4$ ice is dissolved in $N_2$ ice (Cruikshank et al. 1993; Quirico & Schmitt 1997a; Quirico et al. 1999). Triton's water ice absorption was reported to be slightly stronger on the leading hemisphere, while little if any longitudinal variation was seen in the $CO_2$ ice bands, implying a globally homogeneous or high latitude distribution of that species. Now, with our much larger set of spectra from 53 separate nights, we can test the validity of the longitudinal patterns reported earlier as well as searching for additional, more subtle trends not detected in the previous, sparser data set.

The large amplitude longitudinal variation in Triton's $N_2$ ice reported in Paper 1 is confirmed by the new data, as shown in Fig. 3. A sinusoid fitted to the integrated absorption data has its maximum at 31±3° East sub-Earth longitude and a peak-to-peak amplitude of 75±4 percent (comparable to, but smaller than the 96±16 percent variation reported in Paper 1). We measure integrated absorption by normalizing each spectrum to a line fitted to continuum wavelengths on either side of the absorption band (2.093 to 2.117 µm and 2.175 to 1.185 µm, in the case of this $N_2$ ice band), then integrating one minus the normalized spectrum over the band interval (2.117 to 2.175 µm, for this band). We also note that the width and shape of Triton's $N_2$ band are consistent with the hexagonal β phase of $N_2$ ice which is stable above 35.6 K, and not the lower temperature, cubic, α phase (Scott 1976). Additionally, the presence of the shoulder at about 2.16 µm indicates that the temperature of the $N_2$ ice is below about 41 K (see Paper 1; also Grundy et al. 1993; Tryka et al. 1993).

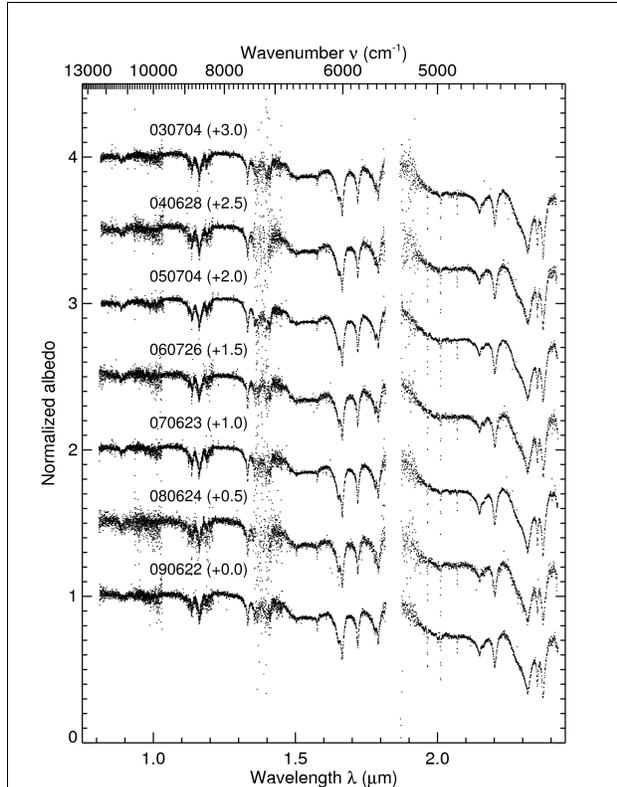

**Fig. 2:** Example spectra from individual nights of observations, labeled according to their UT dates (YYMMDD; see Table 1 for detailed circumstances). The spectra were all normalized, then offset vertically by values shown in parentheses. The 060726 and 070623 spectra provide an interesting comparison, having been obtained at longitudes close to minimum and maximum $N_2$ ice absorption, respectively. The 080624 and 090622 spectra show that useful data quality can be achieved, even with less than an hour of integration on Triton.

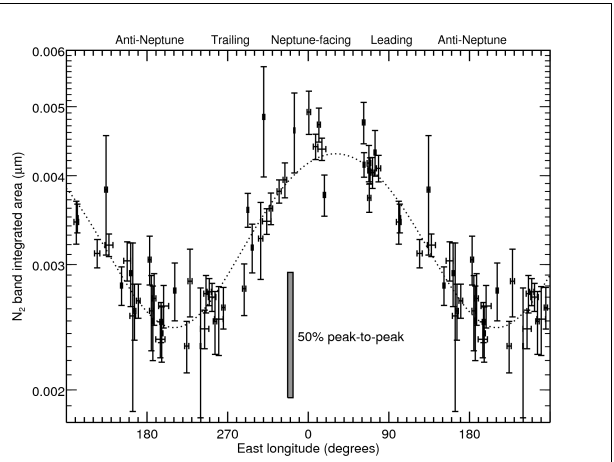

**Fig 3:** Integrated area of the 2.15 µm $N_2$ ice absorption band as a function of sub-Earth longitude on Triton, showing a large periodic variation as Triton rotates. A logarithmic scale is used to facilitate comparison with similar plots to be shown for other ice species. Each point represents observations from a single night, with points duplicated over an additional ±90° to better show the cyclic pattern. The dotted curve is a sinusoid fitted to the data.

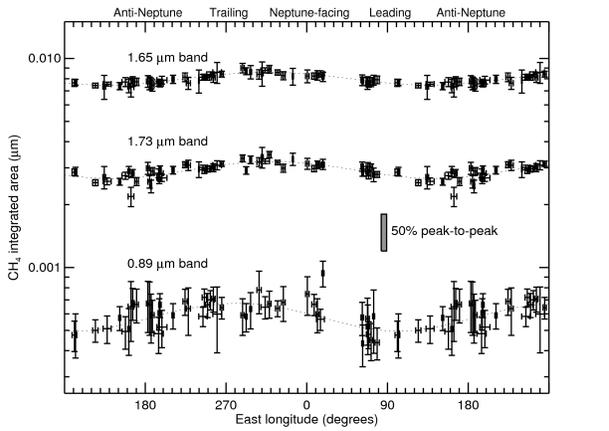

**Fig. 4:** Integrated areas of three $CH_4$ ice absorption bands as a function of sub-Earth longitude on Triton, showing periodic variations as Triton rotates. Other features are as described in Fig. 3.

Triton's methane ice produces numerous absorption bands, which combine to form a series of band complexes diminishing in strength toward shorter wavelengths (see Fig. 1). Paper 1 reported longitudinal variations for three $CH_4$ bands, at 0.89, 1.65, and 1.73 µm, showing maximum absorption on the trailing hemisphere. The new data confirm the earlier results for those three bands (see Fig. 4) and also show similar behavior for the 1.15, 1.33, and 2.2 µm $CH_4$ ice bands (which are omitted from Fig. 4 to reduce clutter). Wavelengths used in computing integrated absorptions throughout this paper are listed in Table 2. From sinusoids fitted to the integrated absorptions, maximum absorption for the six $CH_4$ bands are all near 300 East longitude, corresponding to the Neptune-facing part of the trailing hemisphere. Parameters of these fits are listed in Table 3. Since Triton's $CH_4$ is mostly dissolved in $N_2$ ice (Cruikshank et al. 1993; Quirico & Schmitt 1997a; Quirico et al. 1999), one might expect the two ices to share the same longitudinal distribution. Clearly, the new data confirm that not to be the case. Triton's $CH_4$ and $N_2$ ices evidently do not simply co-occur with a constant mixing ratio. Their mixing ratio must be a function of longitude. Additional clues to the nature of this regionally variable dilution can also be gleaned from the wavelengths of the $CH_4$ absorptions, to be described later.

**Table 2. Wavelengths (µm) used for computing integrated areas of absorption bands**

| Absorption band | Continuum | Band | Continuum |
|---|---|---|---|
| $N_2$ 2.15 µm | 2.093 - 2.117 | 2.117 - 2.175 | 2.175 - 2.185 |
| $CH_4$ 0.89 µm | 0.860 - 0.878 | 0.878 - 0.909 | 0.909 - 0.930 |
| $CH_4$ 1.15 µm | 1.070 - 1.100 | 1.100 - 1.210 | 1.210 - 1.230 |
| $CH_4$ 1.33 µm | 1.240 - 1.290 | 1.290 - 1.342 | 1.342 - 1.343 |
| $CH_4$ 1.65 µm | 1.605 - 1.620 | 1.620 - 1.685 | 1.685 - 1.699 |
| $CH_4$ 1.73 µm | 1.686 - 1.698 | 1.698 - 1.735 | 1.735 - 1.745 |
| $CH_4$ 2.20 µm | 2.170 - 2.180 | 2.180 - 2.225 | 2.225 - 2.235 |
| $CO_2$ 1.965 µm | 1.957 - 1.962 | 1.962 - 1.969 | 1.969 - 1.974 |
| $CO_2$ 2.01 µm | 2.002 - 2.008 | 2.008 - 2.015 | 2.015 - 2.020 |
| $CO_2$ 2.07 µm | 2.062 - 2.068 | 2.068 - 2.072 | 2.072 - 2.078 |
| CO 2.35 µm | 2.336 - 2.345 | 2.345 - 2.358 | 2.358 - 2.367 |

**Table 3. Sinusoidal fits to longitudinal variations of Triton's absorption bands**

| Absorption band | Peak-to-peak amplitude (%) | East longitude of maximum (°) |
|---|---|---|
| $N_2$ 2.15 µm | 75 ± 4 | 31 ± 3 |
| CO 2.35 µm | 72 ± 6 | 56 ± 4 |
| $CH_4$ 0.89 µm | 36 ± 6 | 282 ± 9 |
| $CH_4$ 1.15 µm | 33 ± 7 | 305 ± 12 |
| $CH_4$ 1.33 µm | 28 ± 4 | 288 ± 9 |
| $CH_4$ 1.65 µm | 15 ± 1 | 317 ± 3 |
| $CH_4$ 1.73 µm | 20 ± 2 | 322 ± 5 |
| $CH_4$ 2.20 µm | 19 ± 2 | 299 ± 7 |
| $H_2O$ ratio | 4 ± 2 | 169 ± 26 |
| $CO_2$ triplet | 2 ± 2 | 146 ± 22 |

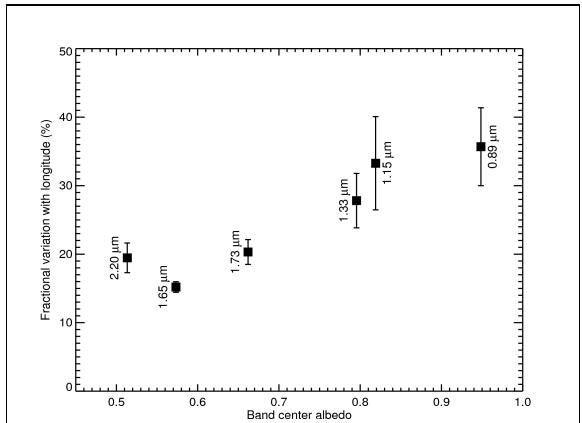

**Fig. 5:** Fractional longitudinal variation of six $CH_4$ ice absorption bands (labeled according to their wavelengths) as a function of the minimum albedo in the band center. Weaker bands, which sample deeper within the surface of Triton, show greater longitudinal variation. The 1.65 µm $CH_4$ band coincides with a water ice band, reducing its apparent variability (as will be discussed shortly, Triton's $H_2O$ absorption shows little longitudinal variation).

Triton's weaker, shorter-wavelength $CH_4$ bands at 0.89, 1.15, and 1.33 µm exhibit stronger fractional variation with longitude (36±6%, 33±7%, and 28±4% peak-to-peak, respectively) than the stronger, longer-wavelength bands at 1.65, 1.73, and 2.2 µm (15±1%, 20±2%, and 19±2%, respectively), as shown in Fig. 5. In general, observations of weaker absorption bands sample more deeply within a surface (e.g., Grundy & Fink 1996; Tegler et al. 2008; Abernathy et al. 2009; Merlin et al. 2009). Since photons at weakly absorbing wavelengths are able to pass through more material before being absorbed, these photons have a better chance of reaching greater depths, and in turn, a better chance of escaping from those depths. Only photons that escape can be detected by remote observers. A trend of greater longitudinal variation for shallower bands suggests that regions with thicker deposits and/or higher concentrations of $CH_4$ ice capable of producing appreciable absorption in the weaker bands are more heterogeneously distributed than the relatively shallow and/or highly diluted layer of $CH_4$-containing ice responsible for the stronger $CH_4$ absorptions. Perhaps most of the observed longitudinal variation in Triton's $CH_4$ bands comes from relatively isolated, low-latitude regions of deep or $CH_4$-rich ice. These regions would need to be preferentially located on the trailing hemisphere, unlike the $N_2$ ice which shows more absorption on the leading part of the sub-Neptune hemisphere.

Our data confirm that Triton's $CH_4$ bands are blue-shifted relative to the wavelengths of pure $CH_4$ ice absorption bands (Grundy et al. 2002a), indicating that Triton's $CH_4$ is diluted in $N_2$ ice (e.g., Quirico & Schmitt 1997a; Quirico et al. 1999). Average blue shifts for three $CH_4$ absorption bands relative to the shifts reported for highly diluted $CH_4$ by Quirico & Schmitt (1997a) are shown as a function of sub-observer longitude in Fig. 6, as measured by autocorrelation. In this plot, a value of zero would correspond to pure $CH_4$ ice while a value of one indicates ice in which each $CH_4$ molecule has exclusively $N_2$ molecules for neighbors. Intermediate dilutions lie between zero and one, but a value of 0.9 should not be interpreted as implying a 90% $N_2$ and 10% $CH_4$ composition, since the dependence of the shift on concentration is not yet well understood (Brunetto et al. 2008; Cornelison et al. 2008). $CH_4$ is not even soluble in $N_2$ ice at 10% concentration, so at least two distinct phases would have to be present at that composition (Prokhvatilov & Yantsevich 1983; Lunine & Stephenson 1985). Indeed, the presence of small areas of $CH_4$-dominated ice is not ruled out by these shifts, with important implications for Triton's atmospheric composition, as will be discussed later. A sinusoidal fit to these measurements shows shifts ranging from 88% to 94% of that of $N_2$-isolated $CH_4$. The peak-to-peak variation of this sinusoid is 0.068±0.011 in these units, a 6-$\sigma$ detection of longitudinal variation. The least diluted $CH_4$ coincides with trailing-hemisphere longitudes where $CH_4$ absorption bands are strongest, implying that higher $CH_4$ concentration in $N_2$ is at least partially responsible for enhancing $CH_4$ absorption at those longitudes.

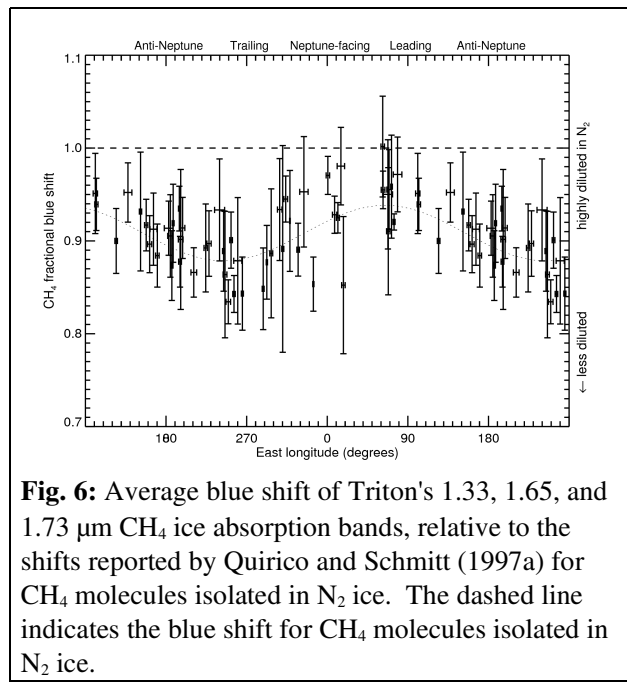

**Fig. 6:** Average blue shift of Triton's 1.33, 1.65, and 1.73 µm $CH_4$ ice absorption bands, relative to the shifts reported by Quirico and Schmitt (1997a) for $CH_4$ molecules isolated in $N_2$ ice. The dashed line indicates the blue shift for $CH_4$ molecules isolated in $N_2$ ice.

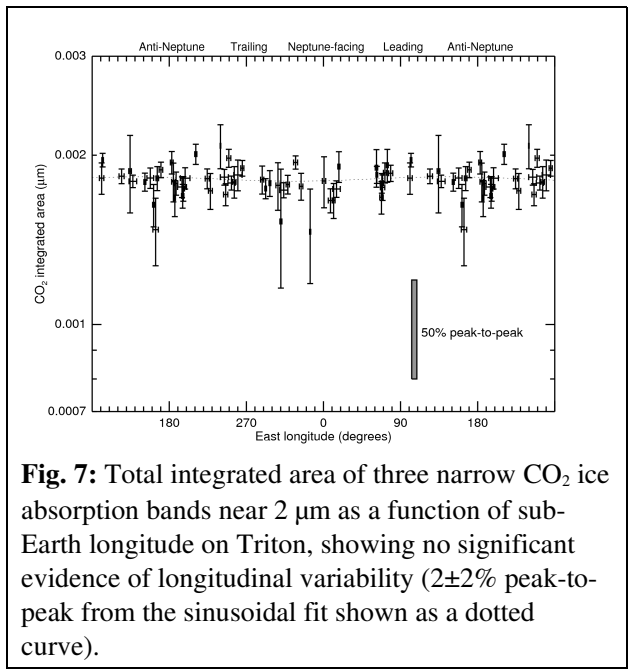

**Fig. 7:** Total integrated area of three narrow $CO_2$ ice absorption bands near 2 µm as a function of sub-Earth longitude on Triton, showing no significant evidence of longitudinal variability (2±2% peak-to-peak from the sinusoidal fit shown as a dotted curve).

Paper 1 reported no measurable longitudinal variation of Triton's $CO_2$ absorption bands around 2 µm. The new data confirm that result, as shown in Fig. 7. To produce such remarkably little variation as Triton rotates, the $CO_2$ ice must be very uniformly distributed and/or be confined to high southern latitudes.

The longitudinal variation of Triton's $H_2O$ ice is trickier to quantify. Although the distinctive shapes of crystalline $H_2O$ ice bands at 1.5 and 2 µm are obvious in the spectra (see Fig. 1 and discussion in Paper 1), these absorption complexes are so broad that they straddle gaps in wavelength coverage as well as overlapping other absorption bands, preventing numerical integration of their areas. Still, it is possible to construct albedo ratios that are sensitive to $H_2O$ ice absorption. In Paper 1 we used a ratio between the center of the 1.5 µm $H_2O$ ice absorption complex and continuum wavelengths on either side. Unfortunately, the longer wavelength continuum region used in that paper (1.683 to 1.710 µm) is contaminated by $CH_4$ absorption, leading to a longitudinal modulation of the continuum, and thus the apparent modulation in $H_2O$ ice absorption reported in Paper 1. To avoid this source of possible contamination in the present analysis, we compute one minus the ratio between the mean albedo from 1.505 to 1.57 µm (the core of the 1.5 µm $H_2O$ absorption) and the mean albedo from 1.22 to 1.27 µm (the continuum). This quantity, shown in Fig. 8, measures the fractional band depth of the 1.5 µm $H_2O$ ice band complex, and thus the ice abundance (assuming no regional variation in ice texture). Contrary to the conclusions of Paper 1, this analysis reveals no significant longitudinal variation in $H_2O$ ice absorption, behavior very similar to that exhibited by the $CO_2$ ice absorptions.

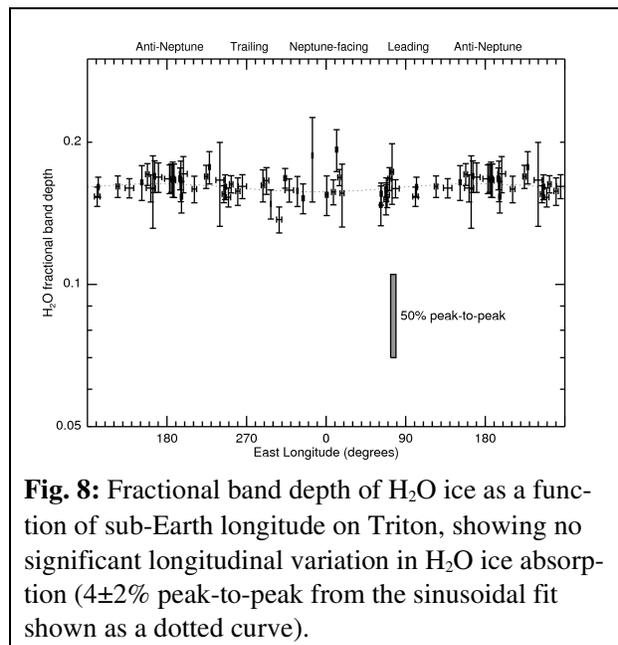

**Fig. 8:** Fractional band depth of $H_2O$ ice as a function of sub-Earth longitude on Triton, showing no significant longitudinal variation in $H_2O$ ice absorption (4±2% peak-to-peak from the sinusoidal fit shown as a dotted curve).

Finally, we turn to carbon monoxide ice. CO has two narrow absorption bands in our wavelength region (Legay & Legay-Sommaire 1982; Quirico & Schmitt 1997b). The 0-3 over-tone at 1.578 µm coincides with a narrow $CO_2$ band and with the corner of the square-bottomed crystalline $H_2O$ ice band, making it effectively unquantifiable in our data. The stronger, 0-2 overtone at 2.352 µm falls between two particularly deep $CH_4$ absorption bands (see Fig. 1). We can compute an integrated area for this band using 2.345 to 2.358 µm for the band and 2.336 to 2.345 µm and 2.358 to 2.367 µm for the continuum, as shown in Fig. 9. The resulting area is likely to be somewhat contaminated by $CH_4$ absorption, and is correspondingly sensitive to the choice of wavelength limits, but as the measured longitudinal variation is much stronger than seen for $CH_4$ absorptions (especially the stronger $CH_4$ bands), we believe this pattern is mostly due to actual variation in CO absorption. A si-nusoidal fit gives a peak-to-peak variation of 72±6% with the maximum on the leading side of the sub-Neptune hemisphere. This distribution looks very similar to what was seen for $N_2$ absorption (see Fig. 3), suggesting that, unlike $CH_4$ ice, CO ice is globally accompanied by $N_2$ ice. CO and $N_2$ ices have very similar vapor pressures (Brown & Ziegler 1980) and they are miscible in one another in all proportions (Scott 1976), so it is reasonable to expect $N_2$ and CO to co-occur on Triton's surface, in contrast with $CH_4$ ice, which is much less volatile. However, we note that the peaks of the sinusoidal fits to the $N_2$ and CO absorption variations are shifted from one another, by amounts that seem to be statistically significant (31±3 and 56±4 degrees longitude for $N_2$ and CO, respectively). That the CO absorption should be shifted slightly relative to the $N_2$ absorption is puzzling in light of their expected co-occurrence, but it could simply be due to contamination of our CO absorption measurement by the adjacent strong $CH_4$ bands.

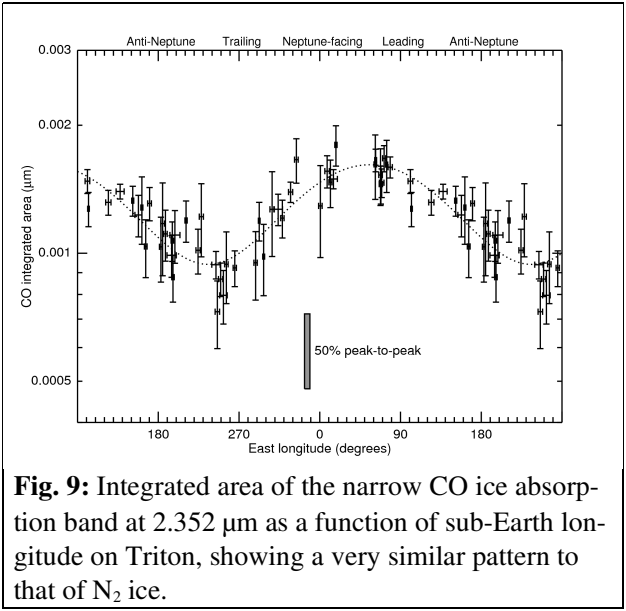

**Fig. 9:** Integrated area of the narrow CO ice absorption band at 2.352 µm as a function of sub-Earth longitude on Triton, showing a very similar pattern to that of $N_2$ ice.

To summarize the observed longitudinal variations, we compare in Fig. 10 the different patterns of variation seen for absorption bands of the five ice species as Triton rotates about its axis. The $N_2$ and CO absorptions show the greatest magnitude of variation, and their variations appear quite similar to one another, implying similar geographic distributions. The absorption variations for six different methane bands are similar to one another, so this plot includes two representative examples: the strongly absorbing 1.73 µm band and the weakly absorbing 0.89 µm band. Both show more absorption on Triton's trailing hemisphere. $H_2O$ and $CO_2$ ice absorptions show remarkably little variation as Triton rotates, similar to the very limited variation in visible wavelength reflectance as measured by a clear filter lightcurve from disk-integrated Voyager approach images obtained at a -50° sub-observer latitude (Smith et al. 1989; Hillier et al. 1991).

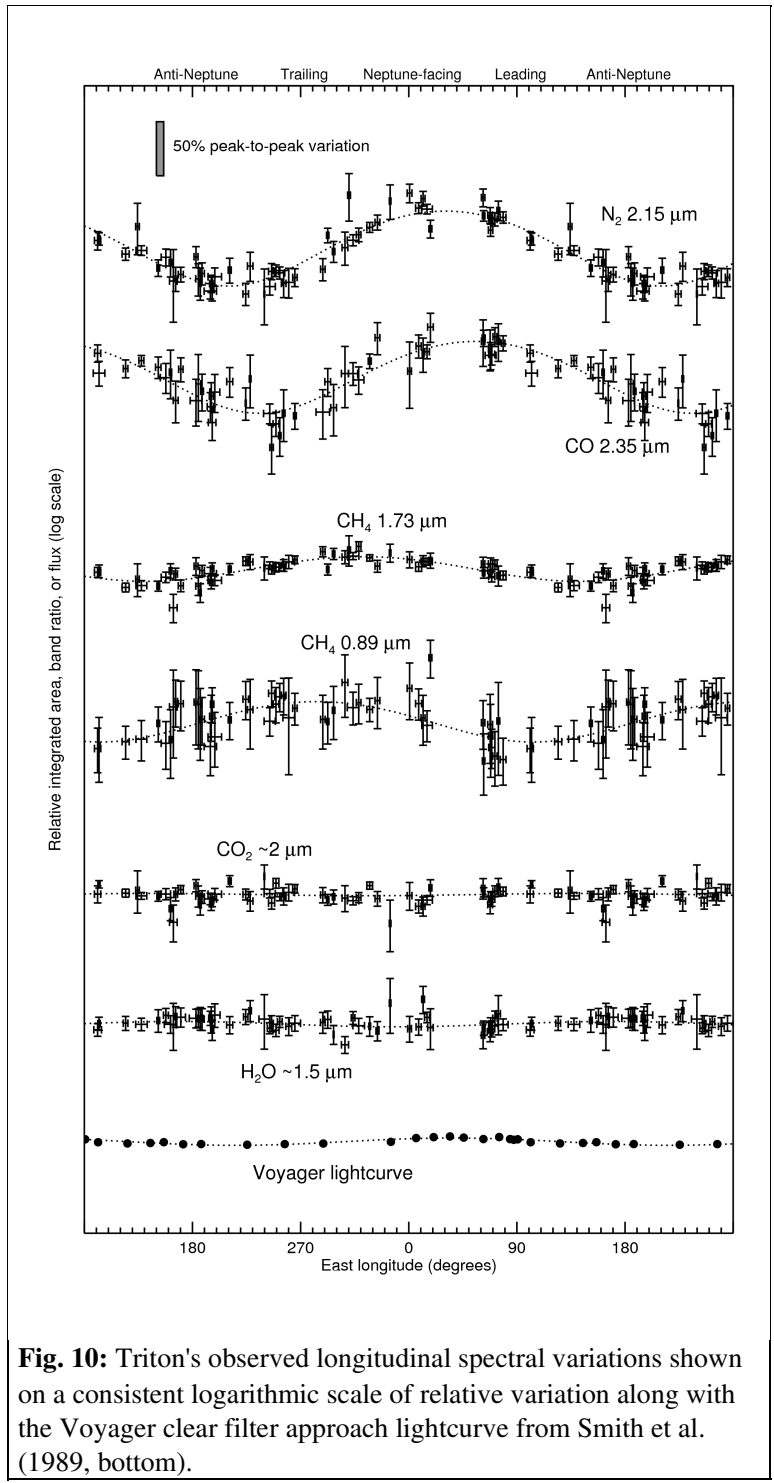

**Fig. 10:** Triton's observed longitudinal spectral variations shown on a consistent logarithmic scale of relative variation along with the Voyager clear filter approach lightcurve from Smith et al. (1989, bottom).

## 4. Search for Secular Change

Triton has been experiencing a major southern summer in recent decades, with the sub-solar latitude reaching a summer solstice maximum of 50° South latitude during the years 2000-2001. Since then, the sub-solar latitude has begun slowly moving back toward the equator (see Table 1, noting that sub-Earth and sub-solar points differ by the phase angle). Continuous sunlight falling on $N_2$ and CO ices at high southern latitudes should cause the ice to sublimate at a rate of the order of centimeters per Earth year (e.g., Trafton et al. 1998). This gradual removal of volatile ice could result in seasonal changes in the observed absorption bands by exposing underlaying strata with distinct compositions, or by driving evolution of the composition or texture of the ice (Eluszkiewicz 1991; Grundy & Stansberry 2000). However, any such changes occurring during the past decade of our observations must be subtle, since the patterns of longitudinal variability for all five ice species exhibit no obvious changes over that interval.

For a more sensitive search for evidence of ice recession, we split our 53 nights of data into two groups, 28 nights collected during 2000-2004 and 25 collected during 2005-2009. Separate sinusoidal fits to the $N_2$ band measurements for these two groups are shown in Fig. 11. The most notable difference is that the later data show less scatter than the earlier data, having benefited from more consistent and polished acquisition techniques, as well as a number of measures taken by IRTF staff over the years to improve image quality. The earlier data generally fall above the dashed curve fitted to the entire data set, especially around the higher parts of that curve near zero longitude (Triton's Neptune-facing hemisphere). The maxima of the two sinusoidal fits to the data subsets are 4.39±0.07 nm and 4.21±0.06 nm for 2000-2004 and 2005-2009, respectively, amounting to a tentative detection of the loss of $N_2$ ice from

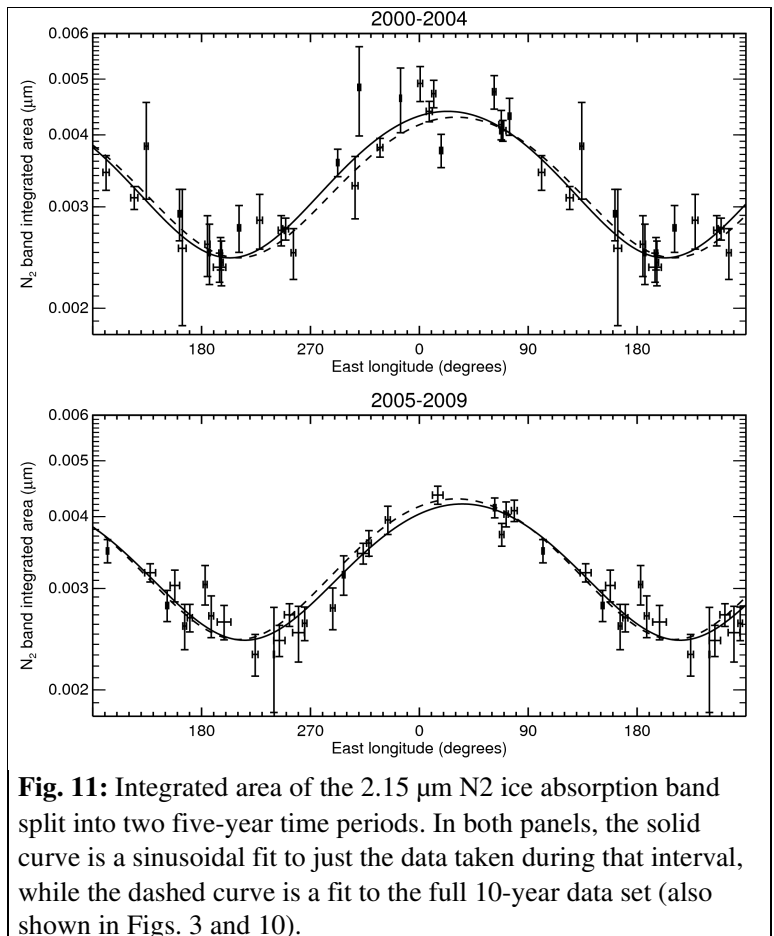

**Fig. 11:** Integrated area of the 2.15 μm N2 ice absorption band split into two five-year time periods. In both panels, the solid curve is a sinusoidal fit to just the data taken during that interval, while the dashed curve is a fit to the full 10-year data set (also shown in Figs. 3 and 10).

Triton's summer hemisphere due to seasonal volatile transport. However, this evidence for seasonal $N_2$ ice sublimation clearly needs testing with additional observations. Its verification will be complicated by the change of Triton's sub-solar latitude, which will be moving toward the equator at an accelerating pace over the coming years. That changing geometry will add an additional source of secular spectral changes likely to be difficult to disentangle from changes caused by volatile transport.

We have seen no convincing evidence for transient changes in our spectra, as have been reported at visible wavelengths (Buratti et al. 1999; Hicks & Buratti 2004). Outlier points do appear in our plots, but they are not more abundant than would be expected from our error statistics. Additionally, the outliers tend to correspond to nights with poor sky conditions or when something went wrong (such as observer error, or failures of equipment, software, power, or the network), preventing us from collecting as much Triton or solar analog data as we had intended.

## 5. Discussion

Here we explore what the observed longitudinal patterns and secular evolution of the ice absorptions can tell us about the spatial distributions and temporal behaviors of Triton's surface ices. We hope to combine our spectral constraints with geological and photometric data from the Voyager encounter to begin to identify the compositions of enigmatic landforms seen in Voyager images, and thus advance our understanding of Triton's seasonal cycles and surface-atmosphere interaction.

We begin with the most volatile ices $N_2$ and CO, which are expected to migrate around Triton's surface on seasonal timescales. Previous investigators computed suites of volatile transport models for various values of unknown parameters such as the thermal inertia, emissivity, and inventory of $N_2$ ice, and compared these with Voyager albedo and color maps as well as the atmospheric pressure. This work was hampered by the fact that Voyager's instruments could not distinguish between Triton's ice species, so it was hard to say whether or not model distributions of $N_2$ ice were consistent with the imagery. Without spectroscopic evidence to the contrary, the large southern polar cap has often been interpreted as consisting mainly of $N_2$ ice (e.g., Croft et al. 1995) despite the preference among volatile transport models for $N_2$ ice being restricted to lower latitudes receiving less insolation (e.g., Stansberry 1989; Spencer 1990; Hansen & Paige 1992; Spencer & Moore 1992; Stansberry et al. 1992; Yelle 1992; Trafton et al. 1998). We take a different approach, combining our observations of spectral variations as Triton rotates with other constraints in an attempt to learn where those ices are presently located. Although the observed spectral variations could conceivably be caused by a wide variety of mechanisms, such as longitudinal variations in texture or stratigraphy, the simplest hypothesis is that they reflect the projected area of geological units composed of the different ices.

In Paper 1, we considered the six McEwen (1990) color and albedo units as possible locations of Triton's $N_2$ ice. We found that no combination of these units had the same longitudinal variation as the $N_2$ band depth. However, if $N_2$ ice were confined to portions of those units North of 31° South latitude, combinations could be found which produced patterns of $N_2$ variation like those observed. To be more quantitative, for each combination of the McEwen units (plus the region unmapped by Voyager, which we call Unit 0) we found the best-fitting cut-off latitude and $N_2$ band integrated area in the $N_2$ absorption-producing region. Combinations having low $\chi_v^2$ values are listed in Table 4. Units 5 and 2 figure in all of the low $\chi_v^2$ lists, and Unit 3 is included in most of them. Unit 5 is the bright South polar cap, Unit 2 is darker streaks superposed on the cap, and Unit 3 is a bright, reddish unit on the polar cap and extending North from the edge of the cap around 300° East longitude.

Table 4.  $N_2$ distributions tied to McEwen (1990) color units

| Units having $N_2$ ice | Latitude cut off (°) | Band area (µm) | Goodness of fit $\chi_v^2$ |
|---|---|---|---|
| 0,2,5 | -31 | 0.017 | 1.13 |
| 2,5 | -34 | 0.015 | 1.30 |
| 2,3,5,6 | -24 | 0.016 | 1.60 |
| 0,2,3,5,6 | -22 | 0.017 | 2.05 |
| 0,2,3,5 | -31 | 0.016 | 2.26 |
| 2,3,4,5,6 | -13 | 0.021 | 2.44 |
| 2,3,5 | -35 | 0.014 | 2.71 |

Table note: Unit 0 represents regions not assigned a unit number by McEwen (1990).

In addition to color and albedo, Triton's surface has been divided into distinct units based on geomorphology and on photometric properties. Lee et al. (1992) identified an "Anomalous-Scattering Region" (ASR) which they interpreted as a "transparent, optically thin, and seasonally controlled veneer of well-annealed solid $N_2$." Plotting the fraction of Triton's visible disk occupied by this unit as a function of sub-observer longitude in Fig. 12, we see that this region is most visible around sub-solar longitude 270°, corresponding to Triton's trailing hemisphere, not 30°, where $N_2$ ice absorption is greatest. Adding in a possible extension of the ASR unit (PE) around 300 East longitude and/or the "Frost Band Region" (FBR) which forms a collar around the polar cap in the Lee et al. map, the maximum projected area remains on the trailing hemisphere, leading us to conclude that the boundaries of these units are unlikely to coincide with the boundaries of Triton's $N_2$ and CO absorbing province. Contrary to the behaviors of $N_2$ and CO, maximum methane absorption is on the trailing hemisphere, suggesting that a higher concentration of $CH_4$, not $N_2$, could be what makes these regions distinct.

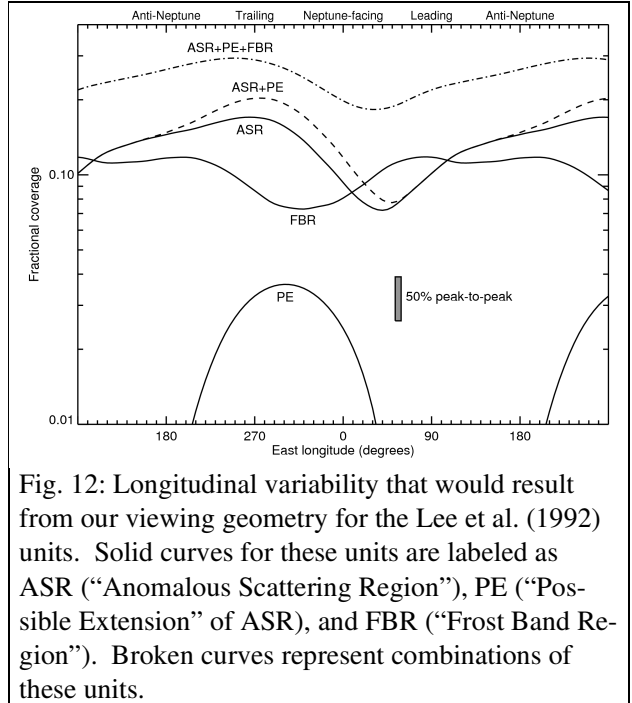

Fig. 12: Longitudinal variability that would result from our viewing geometry for the Lee et al. (1992) units. Solid curves for these units are labeled as ASR ("Anomalous Scattering Region"), PE ("Possible Extension" of ASR), and FBR ("Frost Band Region"). Broken curves represent combinations of these units.

Voyager images of Triton revealed a wealth of landforms, including "cantaloupe terrain", linear ridges, grabens, caldera-like depressions with smooth floors, dark wind streaks, enigmatic dark areas ("guttae") surrounded by bright aureoles, and plains with textures ranging from smooth to hummocky or rugged (Smith et al. 1989; Croft et al. 1995). Unfortunately, geological units have only been mapped systematically on Triton's Neptune-facing hemisphere (but see Stryk & Stooke 2009, for recent progress toward mapping the anti-Neptune hemisphere). The regional nature of the geological maps precludes the same sort of comparison of observed spectral variability with longitudinal visibility of geological units as could be done for the McEwen and Lee et al. units.

We also tried an alternative approach, based on a simple thermal balance model. From the bolometric hemispherical albedo map of Triton (i.e., the spatially resolved analog of the bolometric bond albedo; Stansberry et al. 1992) we calculated a map of diurnal average solar energy absorption as Triton rotates. We reasoned that areas of Triton's surface above some average energy absorption threshold should experience net sublimation of nitrogen ice, the loss of which is balanced by net condensation in regions receiving less energy. We additionally assumed an equilibrium is rapidly reached such that regions receiving more than the threshold energy are denuded of $N_2$ ice while regions receiving less are covered with $N_2$. Alternatively, we could assume that $N_2$ ice remains present in regions experiencing net sublimation but that its texture evolves in such a way that little or no 2.15 µm absorption is seen, whereas condensing regions rapidly acquire a texture that produces an observable absorption band. Regardless, we make the energy threshold and the integrated area of the 2.15 µm $N_2$ absorption band in nitrogen-absorbing areas free parameters and find the values giving the best fit to the observations. The goodness of fit for this model is $\chi_v^2 = 1.33$, for a diurnal average energy threshold of 79.7 erg cm$^{-2}$ s$^{-1}$ and a 2.15 µm $N_2$ band integrated area of 0.012 µm.

This thermal balance model and the truncated McEwen units models both give reasonable matches to our $N_2$ observations, albeit with somewhat different spatial distributions, as shown in which both share a bluish coloration interpreted as resulting from a shallow veneer of freshly-deposited volatile ice (Smith et al. 1989; Lee et al. 1992; Hillier et al. 1994).

Fig. 13. Neither is formally excluded by the data and one could also imagine other models giving comparably good fits. We are forced to conclude that the longitudinal variability plus a model such as one of these is not going to give us a unique map of where Triton's $N_2$ (and by extension CO) absorption occurs. However, both approaches call for high Southern latitudes not to contribute to the observed $N_2$ absorption band, despite being consistently mapped as part of the polar cap (Smith et al. 1989; Thompson & Sagan 1990; Croft et al. 1995). Both models also place $N_2$ ice on a broad swath of the sub-Neptune hemisphere, extending from -30° North toward the equator, and from roughly 280° to 60° East longitude. Geologically, this zone includes some of the northernmost portions of Triton's polar cap, as well as regions of cantaloupe terrain and hummocky plain

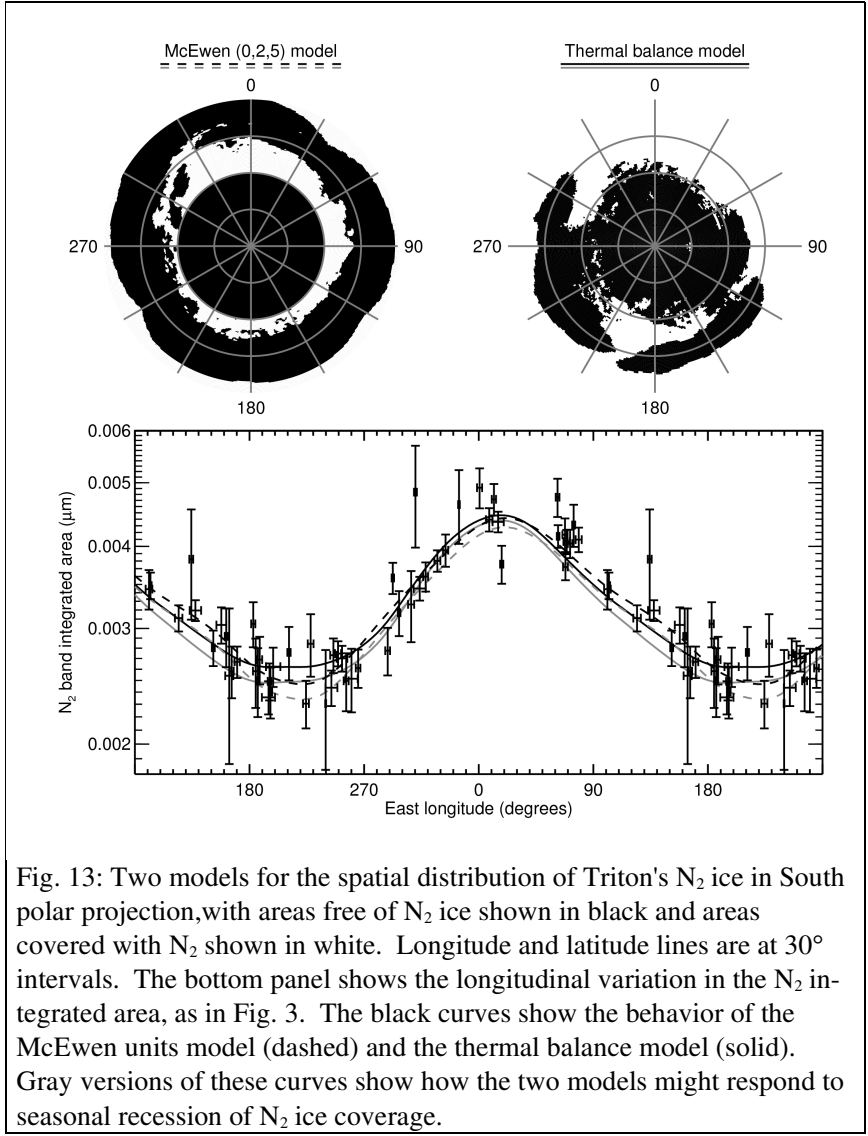

Fig. 13: Two models for the spatial distribution of Triton's $N_2$ ice in South polar projection, with areas free of $N_2$ ice shown in black and areas covered with $N_2$ shown in white. Longitude and latitude lines are at 30° intervals. The bottom panel shows the longitudinal variation in the $N_2$ integrated area, as in Fig. 3. The black curves show the behavior of the McEwen units model (dashed) and the thermal balance model (solid). Gray versions of these curves show how the two models might respond to seasonal recession of $N_2$ ice coverage.

These models call for $N_2$ band integrated areas for $N_2$-covered regions in the range of 0.012-0.021 µm. Are such band areas even plausible for $N_2$ ice? An integrated absorption of 0.02 µm in $N_2$ ice at 38 K (optical constants from Grundy et al. 1993) requires a mean optical path length within $N_2$ ice of about a meter. Such a path length would be produced by a 40 cm thick slab of $N_2$ ice above a scattering substrate, when illuminated and viewed at a mean angle of 30° from the zenith, or by a particulate surface consisting of irregular, but smooth-surfaced 7 cm $N_2$ ice "grains" (using ray-tracing procedures described by Abernathy et al. 2009). Both of these scenarios describe rather unusual surface textures. But such extreme textures seem to be required to provide the mean optical path lengths called for by the observational data, if Triton's $N_2$ absorption band is produced in only a fraction of the total projected surface area, as implied by the observed longitudinal variation.

One possible way to differentiate between $N_2$ ice distribution models such as these is to consider their responses to seasonal recession of $N_2$ ice, in comparison with the secular trends shown in Fig. 11. For the McEwen units model, we simulated seasonal recession by shifting the cut-off latitude Northward slightly, and for the thermal balance model, we did it by decreasing the cut-off energy slightly. The integrated band area in $N_2$-covered areas was held constant. Both simulations resulted in slightly weaker $N_2$ absorption because less area was covered by $N_2$ ice, as shown with gray curves in Fig. 13. The greatest diminution was at longitudes where the $N_2$ absorption was already weakest, leading to increasing fractional variation. This evolutionary pattern differs from that described in Section 4 and shown in Fig. 11. If instead of changing $N_2$ ice areal coverage, seasonal evolution changes the texture of the ice in a way that reduces the band integrated area, $N_2$ absorption would drop most at the longitudes where the most $N_2$ absorption is seen, more consistent with Fig. 11. Textural evolution of the $N_2$ ice is expected from the solar gardening mechanisms described by Grundy & Stansberry (2000), although it is not clear that the trend should be toward the shorter mean optical path lengths in $N_2$ ice necessary to produce a diminishing absorption band.

A completely different approach involves considering the depths of the observed absorption bands. Imagine a region covered with a material that absorbs some fraction of the light from an absorption band, relative to reflectance at a continuum wavelength, surrounded by a region that reflects the same at both wavelengths. The most that the absorbing region can absorb is 100% of the light in the absorption band, so if, in a disk integrated spectrum, that band is 10% deep relative to continuum wavelengths, at least 10% of the observable projected area of the body must have that absorbing material present. This number is a very crude lower limit, since a much larger area could be covered if the absorption in the region in question was less than 100%. The shapes of Triton's absorption bands are consistent with being shallower than 100%, since 100% saturated bands tend to appear broadened, with flat bottoms. A smaller area could be covered if the uncovered area had unrelated absorption at the same wavelength, such as for $CO_2$ and $H_2O$, both of which absorb at 2.01 µm. This situation creates ambiguities in assessing the appropriate continuum level for a band, calling for the exercise of some artistic license. Minimum fractional coverages estimated for various ices using this scheme are tabulated in Table 5, based on the grand average Triton spectrum in Fig. 1.

**Table 5. Estimated minimum fractional coverage for each ice**

| Ice species | Minimum fractional coverage | Absorption band used in estimation |
|---|---|---|
| $N_2$ | 18% | 2.15 µm |
| CO | 27% | 2.35 µm |
| $CH_4$ | 56% | 2.32 µm |
| $CO_2$ | 30% | 2.01 µm |
| $H_2O$ | 28% | 2 µm |

The minimum coverages for all five ices add up to considerably more than 100%, from which we can conclude that some regions must exhibit absorption by more than one ice species. From their similar volatilities and similar longitudinal distributions, we have already proposed that CO and $N_2$ ices co-occur, and from its wavelength shifts, much of Triton's $CH_4$ must be dissolved in $N_2$ ice as well. Combining $N_2$, CO, and $CH_4$ within the 56% of the projected area required for $CH_4$ reduces the minimum total to much closer to 100%. Likewise, it would not be unreasonable to expect the non-volatile species $H_2O$ and $CO_2$, both of which lack significant longitudinal spectral variability, to co-occur in "bedrock" outcrops. These arguments lead to a picture of Triton's surface as being composed of two distinct compositional units, a non-volatile bedrock consisting of $CO_2$ and H2O covering at least 30% of the projected area, plus a volatile veneer composed of $N_2$, CO, and $CH_4$ covering at least 56% of the projected area. Comparable scenarios have been proposed previously (e.g., Quirico et al. 1999), but as described below, our data show that the volatile unit must have regional variations in composition and/or texture.

With its >56% coverage, $CH_4$ ice must be very widely distributed. Its blue-shifted absorption bands imply it is mostly dissolved in $N_2$ ice, yet the longitudinal distribution of the $N_2$ ice absorption implies the nitrogen producing this absorption is distributed over a much smaller fraction of Triton's visible surface. Both $N_2$ distribution models shown in Fig. 13 cover considerably less than 56% of the visible face of Triton. The implication is that some regions that contain both $N_2$ and $CH_4$ ices are not contributing appreciably to the observed 2.15 µm $N_2$ ice absorption band but are contributing $CH_4$ absorption. $N_2$ ice will be spectrally undetectable if its particle size is small enough (mm or smaller), or if it contains abundant small scattering inclusions of another ice with a contrasting refractive index (or even of void space, e.g., Eluszkiewicz and Moncet 2003). Such a situation might result from incorporation of wind-blown dust particles of $H_2O$ or $CO_2$ ice (Grundy et al. 2002b) with refractive indices at these wavelengths of $n = 1.3$ and $n = 1.4$, respectively (Toon et al. 1994; Hansen 1997), or of incomplete or hindered densification (Eluszkiewicz et al. 2007). It could also occur if the temperature-dependent solubility of $CH_4$ in $N_2$ triggers exsolution of tiny methane-rich crystals, with $n = 1.3$ (Pearl et al. 1991). The presence of spectrally invisible $N_2$ ice has important implications for Triton's atmosphere to be discussed in a future paper. More widespread $N_2$ ice would intercept more sunlight and thus could support higher atmospheric pressures. It would also help to explain the presence of geysers discovered by Voyager in Triton's South polar regions and thought to be driven by sublimation of $N_2$ (Smith et al. 1989; Kirk et al. 1990; Soderblom et al. 1990).

An additional complexity of the volatile compositional unit is the observed trend of decreasing longitudinal variability from weak to strong $CH_4$ bands illustrated in Figs. 4 and 5 and Table 3. As discussed earlier, a possible explanation could be the existence of relatively localized regions around 300° longitude having large optical path lengths in $CH_4$. From the shallow depths of the weak $CH_4$ bands (5% for the 0.89 µm band), these regions of enhanced $CH_4$ absorption need not cover much of Triton's surface. They would contribute to the absorption in all of the $CH_4$ bands, but their spectral effect would be most evident in the weakest $CH_4$ bands which are almost too weak to detect in models fitted to the strongest $CH_4$ bands (the stronger bands require far shorter mean optical path lengths in $CH_4$). If these hypothetical regions of greater $CH_4$ path length are comparatively depleted in $N_2$ ice, their $CH_4$ bands will be less blue-shifted than in other regions. This would lead to longitude-dependent shifts in the $CH_4$ bands, with minimum shifts coinciding with the longitudes of these regions. Such a pattern is exactly what we have observed, as plotted in Fig. 6. Unfortunately, we do not have enough information to be able to unambiguously link these anomalously $CH_4$-rich regions to specific areas in Voyager images. However, the Lee et al. (1992) "possible extension of anomalous region" occupies a swath extending from about (280°E,35°S) to about (320°E,0°N), making it a good candidate to be this $CH_4$-rich region. Morphologically, this region resembles cantaloupe terrain, but its appearance is unusual for cantaloupe terrain in featuring lobes of reddish, high albedo material (mapped by McEwen as Unit 3) apparently extending North into a region with lower albedo than most other cantaloupe terrain (mapped by McEwen as Unit 2). It is unclear if this appearance results from new $CH_4$ escaping from Triton's deep interior, from lag deposits produced by preferential sublimation loss of more volatile $N_2$ ice, or from some other process, but additional investigation of this unusual region seems to be warranted.

Neither the presence of $CH_4$ diluted in $N_2$ nor the small areal coverage of the regions of higher $CH_4$ concentration should affect the $N_2$ vapor pressure, as confirmed by the $N_2$ surface pressure of 14 µbars observed by the Voyager 2 radio occultation (Gurrola 1995; Tryka et al.

1993). However, areas of enhanced $CH_4$ (possibly even predominantly $CH_4$) can increase the atmospheric mixing ratio of $CH_4$ over that produced by dilute $CH_4$ alone, as suggested by the atmospheric $CH_4$ mixing ratio measured by the Voyager 2 UV solar occultation (Herbert & Sandel 1991; Cruikshank et al. 1993; Stansberry et al. 1996).

## 6. Conclusion

Near-infrared spectra of Triton show distinct longitudinal variation patterns for absorptions by $N_2$, CO, $CH_4$, $CO_2$, and $H_2O$ ices, constraining the spatial distributions of these different ice species. The variations have persisted over the past decade, implying no wholesale change in the spatial distributions of Triton's ices over that interval.

- Nitrogen ice absorption shows very pronounced oscillation as Triton rotates, being strongest when the leading side of the Neptune-facing hemisphere is oriented toward the observer. This pattern is consistent with a simple thermal balance model. It is also consistent with various subsets of the McEwen (1990) color units, but only if the $N_2$ is confined to parts of those units far from Triton's South pole.
- There is evidence for a subtle seasonal reduction in Triton's $N_2$ ice absorption, but the longitudinal pattern of this decline is consistent with a textural reduction in optical path length in $N_2$ ice, not a retreat of summer hemisphere $N_2$ ice coverage.
- Carbon monoxide absorption is strongest where nitrogen absorption is strongest. The very similar patterns of longitudinal variation, similar volatilities, and miscibility of CO in $N_2$ ice imply co-occurrence.
- Methane ice absorption is distributed very differently from $N_2$ and CO absorption, being strongest on the Neptune-facing side of Triton's trailing hemisphere. Weaker methane ice absorptions probe more deeply into Triton's surface and show stronger longitudinal variation than the stronger methane ice bands do. Wavelength shifts of methane bands indicate that the methane is mostly diluted in nitrogen ice, but the dilution is less where the methane absorptions are strongest. The location of the possible extension of the Lee et al. (1992) anomalous scattering region makes it a reasonable candidate to be this region of higher $CH_4$ concentration.
- Although Triton's methane is mostly diluted in nitrogen ice, the large fraction of Triton's visible area required to produce the strongest $CH_4$ absorption bands is inconsistent with the small fraction of the visible area needed to explain the larger longitudinal variation in the $N_2$ ice band. We conclude that some of Triton's mixed $N_2+CH_4$ ice is not contributing appreciably to the observed 2.15 µm $N_2$ band, perhaps because its texture produces relatively short mean optical path lengths. If CO co-occurs with this spectrally-hidden $N_2$ ice, it too must be hidden in order to match the longitudinal variation of the $N_2$.
- Water and carbon dioxide ice absorptions are remarkably static as Triton rotates, implying globally homogeneous and/or exclusively high-latitude distributions for these non-volatile species.

We intend to continue our near-infrared spectral observations of Triton into the future, in order to better resolve seasonal effects and to continue the search for transient behavior. As one of several small outer solar system bodies with seasonally active volatile ices, these observations will help us understand the growing family which now includes Pluto, Eris, and Makemake. Perhaps others, as-yet undiscovered, will soon be added to their ranks, further boosting the prospects for comparative planetology of these strange, icy worlds.

**Acknowledgments**

We are grateful to W. Golisch, D. Griep, P. Sears, E. Volquardsen, S.J. Bus, and J.T. Rayner for assistance with the IRTF and with SpeX, and to NASA for its support of the IRTF. We thank M.R. Showalter for the Rings Node's on-line ephemeris services, and also NASA's Jet Propulsion Laboratory for the Horizons on-line ephemeris services. We also thank two anonymous reviewers for helpful suggestions and the free and open source communities for empowering us with software used to complete this project, notably Linux, the GNU tools, L*A*T*E*X, FVWM, Tcl/Tk, Python, Evolution, and MySQL. We recognize the cultural significance of the summit of Mauna Kea within the indigenous Hawaiian community and are grateful for the opportunity to observe from this site. This work has been funded by NSF grants AST-0407214 and AST-0085614 and by NASA grants NAG5-4210, NAG5-10497, NAG5-12516, and NNG04G172G.